\documentclass[letterpaper]{article} %
\usepackage[preprint]{aaai2027}  %
\usepackage[hyphens]{url}  %
\usepackage{graphicx} %
\usepackage{natbib}  %
\usepackage{caption} %
\usepackage{algorithm}
\usepackage{algorithmic}

\usepackage{newfloat}
\usepackage{listings}
\DeclareCaptionStyle{ruled}{labelfont=normalfont,labelsep=colon,strut=off} %
\floatstyle{ruled}
\newfloat{listing}{tb}{lst}{}
\floatname{listing}{Listing}

\usepackage{booktabs}

\newcommand{\benchmark}[0]{Decompile-Diverge}
\usepackage{makecell}
\usepackage{xcolor}
\usepackage{pifont}
\usepackage[most]{tcolorbox}
\usepackage{multirow}
\definecolor{cink}{HTML}{24272b}
\definecolor{cmid}{HTML}{2d2f30}

\newcommand{\phead}[2]{%
  \par\noindent
  \vbox{\hsize=\linewidth \offinterlineskip \parindent=0pt \parskip=0pt
    \hbox{\bfseries #1}%
    \kern1.5pt
    \hbox{\small\itshape\color{cmid}#2}%
    \kern1.5pt
    \hbox to\hsize{\color{cmid}\leaders\hrule height0.4pt\hfil}%
  }%
}

\newcommand{\panelw}{0.245\textwidth}
\lstdefinestyle{panel}{language=C,numbers=none,showstringspaces=false,
  basicstyle=\ttfamily\fontsize{8}{9}\selectfont\color{cink},
  keywordstyle={},commentstyle=\color{cmid}\itshape,
  columns=fullflexible,keepspaces=true,aboveskip=1pt,belowskip=0pt,
  escapeinside={<@}{@>}}

\tcbset{cpanel/.style={enhanced,arc=0pt,boxrule=0.6pt,width=\linewidth,
  colframe=cmid,colback=white,left=1pt,right=1pt,top=1pt,bottom=1pt,boxsep=1pt,
  lower separated=true,colbacklower=white,fontlower=\small}}

\definecolor{cerr}{HTML}{C0392B}   %
\definecolor{cok}{HTML}{1E8449}    %

\newcommand{\bad}[1]{\textbf{\color{cerr}#1}}
\newcommand{\okbadge}{\upshape\bfseries\color{cok}Matched}
\newcommand{\divbadge}{\upshape\bfseries\color{cerr}Diverged}
\newcommand{\shipdynamic}[2]{{\small\ttfamily Shipped pass #1 \newline Generated pass #2}}
\newcommand{\io}[3]{%
  {\small\color{cmid} v,\,width,\,height $=$ #1}\\
  {\small\color{cmid}out x,\,y $=$ \bad{#2}\ \newline \ ref x,\,y $=$ #3}}
\usepackage{tikz}

\title{When LLM Decompilers Recompile More and Preserve Less}
\author{
    Chang Liu\textsuperscript{\rm 1},
    Edward Raff\textsuperscript{\rm 2},
    Kristopher Micinski\textsuperscript{\rm 1}
}
\affiliations{
    \textsuperscript{\rm 1}Syracuse University\\
    \textsuperscript{\rm 2}CrowdStrike\\
    cliu57@syr.edu, Edward.Raff@crowdstrike.com, kkmicins@syr.edu
}

\begin{document}

\maketitle

\begin{abstract}

Decompilation recovers high-level source from compiled machine code and
serves as a foundation for security tasks such as vulnerability detection and malware analysis. Traditional decompilers like Ghidra and Hex-Rays expose whatever they cannot resolve as visible placeholders and often emit pseudocode that will not compile or execute; LLM-based decompilers produce clean, idiomatic C and are now judged almost entirely by recompilability and re-executability: whether the output builds and passes its shipped input/output tests. We show that these metrics can reward the wrong path: a function may recompile and pass every shipped test yet diverge on other legitimate inputs, and a disclosed vulnerability may disappear from the recompiled code with no visible trace of the crash. Neither failure is caught by existing suites.

To address this gap, we propose \benchmark{}, a behavioral comparison oracle not relying on fixed or hand-crafted tests: for each function it synthesizes a driver, grows a fuzzing corpus from the reference, and reruns the decompiled code on the same inputs to detect changes in the function's behavior. Across eight systems in nine configurations on established LLM decompilation corpora, candidates that pass every shipped test still diverge from the original on our input corpus: 4.9\% overall, and as many as 13\% for a single system. On 300 real GitHub library functions and 287 CVE-grounded functions, recompilability and behavioral agreement can come apart: the strongest refinement LLM lifts Ghidra's build rate from 75\% to 90\%, while its Matched rate falls from 74\% to 62\%; on disclosed vulnerabilities, up to one tenth exhibit Crash Absence in its output. Source-level analysis traces this divergence to introduced fields, types, callees, and guards that replace the visible unknowns traditional tools leave behind.

\end{abstract}

\section{Introduction}

Decompilation serves as a foundation in both security and software engineering, facilitating critical tasks such as security auditing, vulnerability triage, legacy software maintenance, and malware analysis. As decompiled code is being integrated in active codebases and subjected to fuzzing, behavioral agreement with the original is as critical as readability. Traditional decompilers like Ghidra~\cite{ghidra}, Hex-Rays~\cite{hexrays}, and angr~\cite{angr} often leave certain analysis unresolved and emit non-compilable pseudo code, while the recent LLM-based decompilation systems~\cite{Jiang2023NovaGL,Armengol-Estap'e2023SLaDeAP,Liu2026BinaryDL,Tan2024LLM4DecompileDB,Dramko2025IdiomsND,Tan2025SK2DecompileLT,Hu2024DeGPTOD,Wong2025DecLLMLR} have gained significant improvement in recompilability and re-executability. Generally, they fall into two families: \emph{refinement models}, which refine the output of a traditional decompiler (the \emph{front end}), and \emph{E2E models}, which generate C code directly from assembly.

Recompilability measures whether code can be built into an executable, a prerequisite for re-executability that does not establish it. Re-executability checks a function's output against assertions for given inputs. Many datasets and benchmarks do not provide such checks~\cite{dasilva2021anghabench}. Even when tests are available, their coverage may be too narrow to establish behavioral agreement for decompiled functions. A candidate can therefore recompile and return cleanly on the shipped inputs yet behave differently on other legitimate inputs, as Figure~\ref{fig:decode6} shows.

Motivated by this measurement gap, we propose \benchmark{}. This framework automates evaluation by synthesizing a driver, generating a corpus of fuzzed inputs from the reference function, and comparing a digest of the decompiled code's bounded observable post-state with the corresponding digest from the original. Both LLM-based decompilers and raw front-end baselines are evaluated through this pipeline via two distinct tracks: a general track for behavioral agreement, and a Common Vulnerabilities and Exposures (CVE) track for vulnerability preservation. Our analysis reveals a major shift in failure modes: whereas traditional tools yield visible unknowns, LLMs tend to invent part of the code that either breaks compilation or results in behavioral divergence.
 
Our contributions include:
\begin{itemize}
\item {A behavioral comparison oracle} not relying on fixed tests, for which per-function drivers are synthesized automatically, fuzz inputs are grown dynamically from the reference, and each decompiled output is compared with the reference behavior for each input by checking for crashes, hangs, and differences in the bounded observable post-state.
\item {Two original benchmark sets} that cover 300 real GitHub library functions, and a CVE-grounded track of 287 disclosed vulnerable functions in self-contained suites.
\item {Behavioral agreement can fall as Build rises}, across nine configurations of eight systems. Gains in build rate
coincide with losses in behavioral agreement, and aligned source analysis
traces Divergence and Crash Absence to introduced fields, types, callees, and constants, a common pattern within each system.
\end{itemize}

\section{Background}

\subsection{Traditional Decompilers}

A conventional decompiler lifts machine code to high-level source code, recovers variables and types, restructures control flow, and emits C-like pseudocode~\cite{Cifuentes1994ReverseCT,dream,Basque2024AhoyST,ghidra,hexrays}. Relevant analyses span type and data-structure recovery, exact recovery and recompilation, and learned reconstruction of stripped names~\cite{Noonan2016PolymorphicTI,Zhang2021OSPREYRO,Schulte2018EvolvingED,Lacomis2019DIREAN,Pal2024LenOI,Chen2021AugmentingDO,Xie2024ReSymHL}. Crucially, these systems expose unresolved analysis directly: their pseudocode contains placeholder types such as \texttt{undefined4} and \texttt{\_DWORD}, synthetic names such as \texttt{uVar1} and \texttt{DAT\_*}, raw offsets, and \texttt{goto}-heavy control flow. When compilation fails, diagnostics point to a visible front-end artifact or non-C syntax (Table~\ref{tab:mechanism}). This failure profile provides a baseline for identifying symbols introduced by generative rewriting.

\subsection{LLM-Based Decompilation}

LLMs have become more involved in reverse engineering, and decompilation methodologies have evolved significantly~\cite{Basque2026DecompilingTS}. Early neural systems approached recovery from assembly or LLVM IR as a translation problem~\cite{Katz2019TowardsND,Fu2019ANP,Hosseini2022BeyondTC} and early LLM decompilers rewrote standard decompiler output~\cite{Wong2023RefiningDC}. More recently, two paradigms of LLM decompiler dominate: \emph{end-to-end} systems that map assembly directly to C~\cite{Armengol-Estap'e2023SLaDeAP,Jiang2023NovaGL,Liu2026BinaryDL}, and \emph{refinement} systems that process symbolic pseudocode into idiomatic, compilable C code~\cite{Tan2024LLM4DecompileDB,Dramko2025IdiomsND,Tan2025SK2DecompileLT,Hu2024DeGPTOD}.

\subsection{Decompilation Datasets and Benchmarks}

Learning-based approaches inherently rely on extensive binary corpora. Various datasets supply the critical data necessary to evaluate decompilation generalization, scalability and vulnerability preservation~\cite{Liu2024AssemblageAB,Joyce2025EMBER2024A,Kim2020RevisitingBC,anderson2018emberopendatasettraining,Saul2024IsFS,Dolan-Gavitt2016LAVALA,Hazimeh2020Magma,Mei2024ARVOAO,Zhang2018PreciseAA}, while large scale datasets provide more realistic evaluation~\cite{Tan2025DecompileBenchMB}. However, LLM decompilers are predominantly evaluated on highly constrained benchmarks: HumanEval-Decompile, MBPP C conversion, ExeBench, and AnghaBench~\cite{Tan2024LLM4DecompileDB,Tan2025SK2DecompileLT,armengol2022exebench,dasilva2021anghabench}. These datasets primarily offer short functions, flat scalar signatures, and sparse input-output assertions. Consequently, claims of behavioral agreement rely heavily on recompilability and pass@$k$ re-executability~\cite{Chen2021EvaluatingLL} over the fixed tests.

Static tests and superficial similarity fail to establish behavioral agreement~\cite{Liu2023IsYC,Tan2024LLM4DecompileDB,Cao2024EvaluatingTE,Dramko2024ATO}. State-of-the-art evaluations remain limited: they often restrict LLMs to projects with pre-existing fuzzing suites~\cite{Gao2025DecompileBenchAC} and report only fuzzing coverage, omitting the broader analytic techniques used on conventional decompilers~\cite{Liu2020HowFW,Zou2024DHelixAG}. Csmith~\cite{Yang2011FindingAU} generates random C programs and compares compilers on them; equivalence modulo inputs~\cite{Le2014CompilerVV} derives program variants that must agree on a profiled input set. Decompiler testing adopted this program-side recipe: DecFuzzer~\cite{Liu2020HowFW} recompiles decompiled Csmith and programs and compares executions; Bin2Wrong~\cite{bin2wrong} mutates source, compiler, optimization, and executable format as one testcase, and D-Helix~\cite{Zou2024DHelixAG} compares original and recompiled binaries by symbolic differentiation. All vary the program and target conventional decompilers, so their subjects carry none of the project-specific structs, typedefs, or wrappers that LLM refiners replace (Section~\ref{sec:source-analysis}). CHISEL~\cite{kohli2026chiselingsourcecodeaienabled} instead varies the \emph{inputs}, using a coverage-guided fuzzer as in-loop feedback for LLM repair of Ghidra pseudo-C on 120 ExeBench functions.

\section{Motivation}

LLM decompilation systems that finetune on existing models are judged, and increasingly trained, on a surface proxy for behavioral agreement: emit code that compiles and reproduces the reference on the inputs already on hand. LLM4Decompile is fine-tuned with the next-token objective over the reference tokens $y=(y_1,\dots,y_n)$ given the front-end input $x$, 
\begin{equation}
\mathcal{L}(\theta) \;=\; -\sum_{t=1}^{n}
   \log P_{\theta}\!\left(y_t \mid y_{<t},\, x\right),
\label{eq:ce}
\end{equation}
which rewards resemblance to the reference string without measuring behavioral agreement. SK2Decompile uses reinforcement learning with rewards that inspect the same surface. Its structure reward is zero when the IR does not compile and otherwise adds the Jaccard overlap of placeholder identifiers
$r_{\mathrm{ph}}=|I_{\mathrm{gen}}\cap I_{\mathrm{IR}}|/|I_{\mathrm{gen}}\cup I_{\mathrm{IR}}|$ to a base reward of 1.0:
\begin{equation}
r_{\mathrm{struct}} =
\left\{\!
\begin{array}{ll}
0.0, & \mbox{if IR does not compile,}\\
1.0 + r_{\mathrm{ph}}, & \mbox{if IR compiles,}
\end{array}
\right.
\label{eq:sk2}
\end{equation}
A second phase rewards identifier-name similarity. No term of either objective asks whether the recovered function agrees with the reference on an input it was never shown, and re-executability is only reported on shipped tests. A rewrite that compiles, reads cleanly, and passes every released input output pair thus receives full credit even when the function's behavior changes, which prior human studies also confirm~\cite{Basque2026DecompilingTS,Votipka2019AnOI,Burk2022DecompersonHH}. Recent work adds runtime feedback, reward shaping, and structural guidance~\cite{Wong2025DecLLMLR,Zou2025DLiFTIL,Wang2025SALT4DecompileIS,Shypula2026DecafIN} but leaves the proxy in place.

\begin{figure*}[t]
\noindent
\begin{minipage}[t]{\panelw}\vspace{0pt}%
\begin{tcolorbox}[cpanel]
\phead{original}{reference}
\begin{lstlisting}[style=panel]
void decode6(long v,
  long width, long height,
  long *x, long *y){
 long w, D = 6;
 long col = v/(height*6);
 long maxcol = (width-4)/6;
<@\hl@> if( col >= maxcol ){
  col = maxcol;
  D = width - maxcol*6; }
 w = v - col*height*6;
<@\hl@> *x = col*6 + w%D;
 *y = height-1 - w/D;
}
\end{lstlisting}
\tcblower
{\small\ttfamily shipped: 10 IO pairs}\\
{\small\color{cmid}args $\in[0,124]$; none reaches the final band}
\end{tcolorbox}
\end{minipage}\hfill
\begin{minipage}[t]{\panelw}\vspace{0pt}%
\begin{tcolorbox}[cpanel]
\phead{raw Hex-Rays}{\okbadge}
\begin{lstlisting}[style=panel]
void __cdecl decode6(long v,
...
 col = v / (6*height);
 maxcol = (width-4)/6;
<@\hl@> if ( col >= maxcol ){
  col = (width-4)/6;
  D = width - 6*maxcol; }
 *x = (v - 6*height*col)
<@\hl@>      % D + 6*col;
 *y = height - 1
    - (v - 6*height*col)/D;
}
\end{lstlisting}
\tcblower
\shipdynamic{10/10}{100/100}\\
{\small\color{cmid}x,\,y match the reference on every input}
\end{tcolorbox}
\end{minipage}\hfill
\begin{minipage}[t]{\panelw}\vspace{0pt}%
\begin{tcolorbox}[cpanel]
\phead{SK2Decompile}{\divbadge}
\begin{lstlisting}[style=panel]
void decode6(long i,
  long n, long m,
  long *x, long *y){
 long k, l, r;
 r = 6;
 k = i/(m*r);
 l = (n-4)/r;
 if( k >= l ){
  k = l;
<@\hlr@>  r = n - l*r; }
 *x = (i-k*m*r)%r
<@\hlr@>      + k*r;
 *y = m-1 - (i-k*m*r)/r;
}
\end{lstlisting}
\tcblower
\shipdynamic{10/10}{70/100}\\
\io{100,\,16,\,8}{8,\,$-2$}{12,\,6}
\end{tcolorbox}
\end{minipage}\hfill
\begin{minipage}[t]{\panelw}\vspace{0pt}%
\begin{tcolorbox}[cpanel]
\phead{LLM4Decompile}{\divbadge}
\begin{lstlisting}[style=panel]
void decode6(long n,
  long width, long height,
  long *x, long *y){
 long w, D, col;
 w = (width-4)/6;
<@\hlr@> if( n >= w*height*6 ){
  D = width - w*6;
  col = w;
 }else{ D = 6;
  col = n/(height*6); }
 n -= col*height*6;
 *x = col*6 + n%D;
 *y = (height-1) - n/D;
}
\end{lstlisting}
\tcblower
\shipdynamic{10/10}{66/100}\\
\io{0,\,$10^{10}$,\,$10^{9}$}{$\sim$$10^{10}$,\,$-2{\times}10^{18}$}{0,\,$\sim$$10^{9}$}
\end{tcolorbox}
\end{minipage}
\caption{Re-executability does not imply behavioral equivalence.
The three reconstructions are outputs from decompilation systems on the same target, which maps a linear index into bands of width 6 followed by a
variable-width final band. All three agree with the reference on the ten
shipped ExeBench tuples. Inspection of those tuples shows that none
exercises SK2Decompile's final-band error, while their small positive
operands do not overflow LLM4Decompile's threshold product. On the
100-input stress corpus, raw Hex-Rays agrees with the reference on
100/100 inputs, SK2Decompile on 70/100, and LLM4Decompile on 66/100.
Gray highlights the reference check and the $*x$ computation.
SK2Decompile conflates the fixed stride 6 with the mutable final-band
width $D$, then uses $D$ where 6 is required in both the residual and
the horizontal band offset, so both $x$ and $y$ may be corrupted.
LLM4Decompile replaces the quotient check with
$n \ge \mathrm{maxcol}\cdot\mathrm{height}\cdot6$; this comparison is
equivalent only under the intended sign constraints and when the product
does not overflow. It also moves the quotient computation into one
branch, changing division-by-zero fault behavior. Footers show a
representative model output and the corresponding reference output.}
\label{fig:decode6}
\end{figure*}

ExeBench, widely used to evaluate LLM decompilers, has a test function \texttt{decode6}, which maps an index into six-column bands with a narrower final band (Figure~\ref{fig:decode6}). Its ten shipped input–output pairs stay in a narrow range: every argument lies within 0 and 124, and on all ten the quotient col is smaller than maxcol, so the final band is never exercised and the small operands never overflow. SK2Decompile and LLM4Decompile both recompile and pass all ten. However, the dynamically generated input corpus exposes the rewrites: SK2Decompile collapses the fixed stride \texttt{6} and the mutable band width \texttt{D} into a single variable, so on $(v,\text{width},\text{height})=(100,16,8)$ it writes $(x,y)=(8,-2)$ where the reference gives $(12,6)$; it matches only 70/100 inputs. LLM4Decompile multiplies out the division-based band guard, introducing 64-bit overflow: on $(0,10^{10},10^{9})$ it returns roughly $(10^{10},-2{\times}10^{18})$ against the reference's $(0,\sim\!10^{9})$, and matches 66/100. Raw Ghidra and Hex-Rays preserve the reference's operations and match all 100. Other Divergence cases include LLM4Decompile narrowing \texttt{image\_fit}'s unsigned dimensions and changing \texttt{SetFloat}'s constant, SK2Decompile dropping a cast in \texttt{tableset}, and AutoDecompiler emitting uninitialized stack arrays whose in-bounds reads evade AddressSanitizer~\cite{armengol2022exebench,Liu2026BinaryDL,addresssanitizer}. The same pattern recurs elsewhere: replaying every candidate shows that the shipped suites released by the three corpora fail to capture 3\% to 45\% of the divergence in function behavior.

\section{Benchmark Design}

\begin{figure*}[t]
    \centering
    \includegraphics[width=0.97\textwidth]{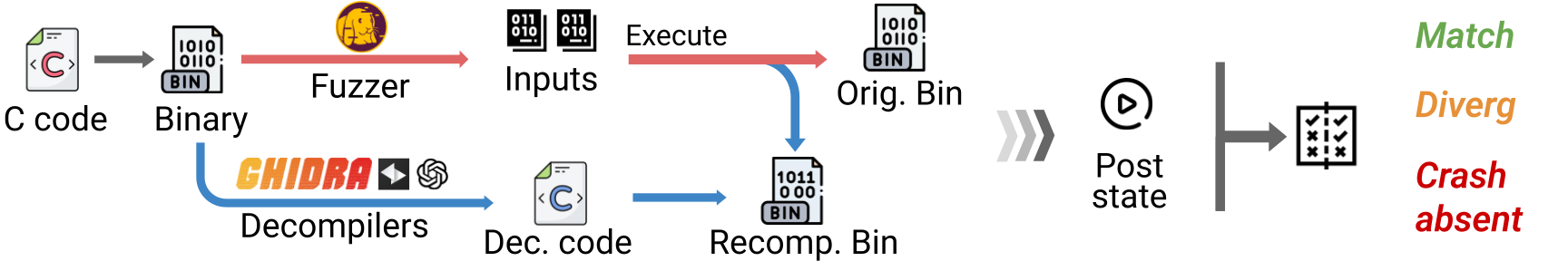}
    \caption{A reference C function is compiled to a binary that a fuzzer explores to build a fixed input corpus. Decompiler recovers C source that is recompiled into a candidate binary. The same inputs are then re-executed on both original and recompiled binaries, and their bounded observable post-states are compared to judge the function Matched (behavior preserved), Diverged (a changed output or bounded observable post-state), or Crash Absence (a reference crash that is silently absent, CVE only).}
    \label{fig:workflow}
\end{figure*}

\subsection{Oracle Design}

A high level workflow of \benchmark{} is illustrated in Figure~\ref{fig:workflow}. It constructs a behavioral comparison oracle for each function from its original source. It synthesizes a driver from the target function, uses AFL++~\cite{aflplusplus} to fuzz the reference implementation, and generates the resulting corpus before evaluating any decompiler outputs. For portability and reproducibility purposes, the portable version provides the inputs we used to run the experiments stated in Section~\ref{sec:exp} and Table~\ref{tab:reach}, and the benchmark has an option to generate the inputs dynamically. The oracle supports most common types including fixed-width scalars, strings, pointer-length arrays, and bounded numeric buffers; Table~\ref{tab:reach} reports its coverage given the time limit stated in Section~\ref{sec:exp}. Each decompiled candidate is then spliced in as-is and executed on exactly the same inputs as the reference. An AddressSanitizer build detects crashes and hangs, while an uninstrumented \texttt{-O0} build compares the bounded observable post-state including the function's return value, the bytes written, process's writable globals, the \texttt{.data} and \texttt{.bss} ranges delimited by linker symbols. The reference and candidate are then compared symbol by symbol through the table, so the check is independent of memory layout (heap, stack, and registers are excluded as relink may change layout). Two further guards keep unstable or environment-dependent behavior out of the judgments: pointer-valued state is relocation-masked in the digest of the bounded observable post-state, and on the GitHub and CVE tracks a divergence counts only if it reproduces on four re-runs of both reference and candidate, discarding flakiness from address-space layout or uninitialized memory. The reference execution thus supplies the expected behavior without hand-written input/output assertions. Because the comparison is limited to the generated corpus and the bounded observable post-state, the reported Matched rate is an upper bound on behavioral agreement across all valid inputs, whereas Divergence and Crash Absence rates are lower bounds. 

\subsection{Benchmark Data Sources}
\label{sec:datas}
\paragraph{Established LLM Decompiler Corpora}

Our setup replaces the shipped input/output tests for four popular corpora used by LLM decompilation systems: HumanEval-Decompile (162)~\cite{Tan2024LLM4DecompileDB}, the ExeBench \textit{valid\_real} split (1{,}551 retained)~\cite{armengol2022exebench}, AnghaBench (218)~\cite{dasilva2021anghabench}, and the MBPP C conversion (603)~\cite{Tan2025SK2DecompileLT}.

\paragraph{GitHub Repositories}
The GitHub track contains 300 C functions (median 23 SLOC) from 132 libraries in 106 repositories without released function-level tests, at most 21 from any one library. Selection is model-blind and restricted to functions that can be exercised through our oracle fuzzer interface; among the 300, seven functions with multi-level pointers, function pointers, multidimensional arrays, by-value structs, or project-specific aggregates use manually written drivers with the same file-input and observation contract. Nine references do not build under the oracle harness and are excluded, leaving the $n{=}291$ of Table~\ref{tab:results-dd}.

\paragraph{Common Vulnerabilities and Exposures}
The Common Vulnerabilities and Exposures (\texttt{CVE}) track contains 287 vulnerable C functions (median 47 SLOC) from 94 open-source projects. We recover 300 functions from vulnerable revisions identified through OSV.dev fix commits~\cite{OSVdev} and from the CVEfixes dataset~\cite{cvefixes}, where 13 are dropped due to fuzzer compatibility and duplication, leaving us 287 in total. Each suite preserves the vulnerable function's original type definitions and helper routines, with minimal stubs for dependencies. Decompilers see the functions' assembly compiled without instrumentation; the testing suites compile these functions with AddressSanitizer and UBSan. A file-input driver passes attacker-controlled bytes through the target function and includes both vulnerability-triggering and safe inputs. AFL++ outputs the replay corpus of 14,323 inputs, including manually constructed proofs of concept when fuzzing cannot recover the crash. The final generated input corpus contains 21,119 inputs, averaging 74 per function; Table~\ref{tab:reach} reports coverage.

\subsection{Systems Benchmarked}

We evaluate eight systems in nine configurations: four refinement systems including LLM4Decompile~\cite{Tan2024LLM4DecompileDB}, Idioms~\cite{Dramko2025IdiomsND}, SK2Decompile~\cite{Tan2025SK2DecompileLT}, and DeGPT~\cite{Hu2024DeGPTOD}; three end-to-end systems including AutoDecompiler~\cite{Liu2026BinaryDL}, Nova~\cite{Jiang2023NovaGL}, and SLaDe~\cite{Armengol-Estap'e2023SLaDeAP}, plus GLM-5.2~\cite{glm5} in both roles. LLM4Decompile and DeGPT refine Ghidra output; the other refinement systems use Hex-Rays. All systems use their published settings, full checkpoints, and prompts. Because DeGPT's original \texttt{gpt-3.5-turbo} backend is limited to 4,096 output tokens and is no longer representative of current chat models, we replace it with Qwen3.6-35B-A3B-FP8~\cite{qwen36_35b_a3b} and label the configuration \emph{DeGPT-Qwen} throughout. These results characterize DeGPT's pipeline with the Qwen backend; the published GPT-3.5-Turbo system is outside our evaluation. SLaDe follows its published source-to-assembly and beam-selection pipeline, using assembly generated from source. Other work~\cite{Wong2025DecLLMLR,Wang2025SALT4DecompileIS,Zou2025DLiFTIL} is excluded because sufficient public artifacts for evaluation were unavailable at the time of experiment.

\section{Evaluation}

\subsection{Experiment Setup and Judgment}
\label{sec:exp}

Let C be the set of successfully built candidates from Section~\ref{sec:datas}, each candidate receives exactly one judgment through the map
\begin{equation}
\begin{aligned}
\mathcal{Y} &= \{\mathrm{Matched},\, \mathrm{Divergence},\, \mathrm{Crash\ Absence}\}, \\
J &: \mathcal{C} \to \mathcal{Y},
\end{aligned}
\end{equation}
where the labels are:
\begin{itemize}
    \item {Matched}: matches the reference on all inputs.
    \item {Divergence}: diverges from the reference on at least one input, changing the observed state including crash, behavioral change, or non-termination.
    \item {Crash Absence}: removes a failure the reference is expected to exhibit, reported only in the CVE track.
\end{itemize}
Since a candidate may satisfy both the Divergence and Crash Absence conditions, we give Crash Absence precedence over Divergence so that $J$ is well defined and the three labels are mutually exclusive.

There are several types the fuzzer will not apply to (e.g., multi-dimensional arrays and function pointers) and several functions will not be built with AddressSanitizer, and these are not included. As a result, the tests are applicable on a total of 3,089 functions, where the final denominators are $n{=}161$, $1{,}538$, $215$, and $597$ for the four established corpora, $291$ for GitHub, $287$ for CVE, in all of which the original references are judged as Matched. We also test fuzzing the established corpora at time budgets of 1 and 5 minutes; the 5-min run expands coverage on only thirteen ExeBench, three AnghaBench, and six MBPP functions, so we stop at the 5 minute fuzzing for existing corpora. For GitHub and CVE functions we set a 10 minutes time limit for fuzzing (further extending the limit does not increase coverage), and the whole fuzzing coverage is shown in Table~\ref{tab:reach}.

\begin{table}[h]
\centering
\small
\setlength{\tabcolsep}{4.5pt}
\begin{tabular}{@{}lcccccc@{}}
\toprule
 & \makecell{Human\\Eval} & \makecell{Exe\\Bench} & \makecell{Angha\\Bench} & \makecell{MBPP} & \makecell{GitHub} & \makecell{CVE} \\
\midrule
Line   & 98 & 99 & 94 & 98 & 90 & 84 \\
Branch & 96 & 93 & 86 & 97 & 82 & 74 \\
\bottomrule
\end{tabular}
\caption{Oracle generated input corpus coverage. \emph{Line} and \emph{Branch} are mean per-function coverage percentages.}
\label{tab:reach}
\end{table}

\subsection{Divergence across Corpora}

\paragraph*{Passing Established-Corpus Tests while Diverging}

\begin{table}[t]
\centering
\setlength{\tabcolsep}{1pt}
\small
\begin{tabular}{@{}l rrr rrr r rrr@{}}
\toprule
& \multicolumn{3}{c}{HumanEval} & \multicolumn{3}{c}{ExeBench} & \multicolumn{1}{c}{Angha} & \multicolumn{3}{c}{MBPP} \\
& \multicolumn{3}{c}{$n{=}161$} & \multicolumn{3}{c}{$n{=}1538$} & \multicolumn{1}{c}{$n{=}215$} & \multicolumn{3}{c}{$n{=}597$} \\
\cmidrule(lr){2-4}\cmidrule(lr){5-7}\cmidrule(lr){8-8}\cmidrule(l){9-11}
System & Pass & Div & {D$|$P} & Pass & Div & {D$|$P} & Div & Pass & Div & {D$|$P} \\
\midrule
\multicolumn{11}{@{}l}{\emph{Raw front ends}} \\
Ghidra & 81 & 1 & 2 & 58 & 4 & 1 & 1 & 73 & 1 & 1 \\
Hex-Rays & 86 & 1 & 1 & 75 & 3 & 5 & 0 & 87 & 1 & 1 \\
\midrule
\multicolumn{11}{@{}l}{\emph{Refinement systems}} \\
LLM4Decompile & 85 & 15 & 7 & 61 & 21 & 8 & 13 & 80 & 14 & 4 \\
SK2Decompile & 50 & 6 & 4 & 69 & 17 & 6 & 25 & 60 & 7 & 4 \\
Idioms & 1 & 0 & - & 6 & 2 & 5 & 0 & 0 & 0 & - \\
GLM-5.2-Refine & 61 & 4 & 1 & 67 & 10 & 6 & 10 & 70 & 7 & 6 \\
DeGPT-Qwen & 81 & 2 & 1 & 59 & 5 & 2 & 2 & 70 & 4 & 4 \\
\midrule
\multicolumn{11}{@{}l}{\emph{End-to-end systems}} \\
AutoDecompiler & 1 & 7 & 0$^{*}$ & 5 & 19 & 9 & 16 & 2 & 14 & 14$^{*}$ \\
Nova & 34 & 55 & 13 & 31 & 30 & 10 & 33 & 41 & 45 & 12 \\
SLaDe & 51 & 11 & 4 & 42 & 13 & 3 & 10 & 53 & 17 & 8 \\
GLM-5.2-E2E & 72 & 25 & 5 & 64 & 17 & 5 & 29 & 76 & 22 & 5 \\
\bottomrule
\end{tabular}
\caption{{Differential testing on the four established corpora.} Pass and Div are percentages of each corpus's denominator $n$. Pass: the candidate builds and matches every input--output pair shipped with the corpus. Div: Divergence, any divergence on at least one input of the generated corpus. Div$|$P: of the candidates in Pass that \benchmark{} also scores, the percentage that diverge anyway; the denominator is that Pass count, not $n$. AnghaBench ships no tests, so only Div is defined there. A dash marks no scorable passer, $^{*}$ fewer than 20 of them.}
\label{tab:results}
\end{table}

On the four established corpora (HumanEval, ExeBench, AnghaBench, MBPP), Table~\ref{tab:results} shows the tradeoff in the terms each corpus itself supplies: refinement clears more of the shipped tests while diverging far more often on the generated corpus. LLM4Decompile passes more than Ghidra on every corpus that ships tests (85 against 81 on HumanEval), yet diverges on 13--21\% of functions against Ghidra's 1--4\%, and SK2Decompile reaches 25\% Divergence on AnghaBench against Hex-Rays's 0\%. The raw front ends stay at or below 4\% Divergence on every corpus, and in paired comparisons 91\% of LLM4Decompile's Divergence cases and 84\% of SK2Decompile's Divergence cases arise only after refinement; only DeGPT-Qwen, which edits in place, stays at $\le$5\% Divergence. Crucially, the shipped tests do not catch this. The Div$|$P column isolates the candidates that pass \emph{every} shipped test and still diverge, reaching 13\% for Nova on HumanEval and 8\% for LLM4Decompile on ExeBench, and pooled it puts every refinement system above its own front end. Replaying every candidate through the three released suites (ported precisely so the shipped tests and \benchmark{} score identical bytes), 4.9\% of the 12{,}133 such passers diverge, 77\% of them by changing an output with no crash at all. Pooled over the 15,379 candidates scored by both, the two oracles disagree on 5.1\%; and 3.9\% pass every shipped test yet diverge on the generated corpus, while 1.2\% fail a shipped test yet stay Matched.

\begin{table}[t]
\centering
\setlength{\tabcolsep}{4pt}
\small
\begin{tabular}{@{}l rrr rrrr@{}}
\toprule
& \multicolumn{3}{c}{GitHub} & \multicolumn{4}{c}{CVE} \\
& \multicolumn{3}{c}{$n{=}291$} & \multicolumn{4}{c}{$n{=}287$} \\
\cmidrule(lr){2-4}\cmidrule(l){5-8}
System & Bld & Mat. & Div & Bld & Mat. & Div & C-A \\
\midrule
\multicolumn{8}{@{}l}{\emph{Raw front ends}} \\
Ghidra & 75 & 74 & 1 & 30 & 27 & 3 & 0 \\
Hex-Rays & 81 & 78 & 3 & 23 & 19 & 3 & 1 \\
\midrule
\multicolumn{8}{@{}l}{\emph{Refinement systems}} \\
LLM4Decompile & 90 & 62 & 28 & 64 & 38 & 17 & 9 \\
SK2Decompile & 41 & 32 & 9 & 11 & 7 & 2 & 2 \\
Idioms & 2 & 1 & 1 & 1 & 1 & 0 & 0 \\
GLM-5.2-Refine & 80 & 71 & 9 & 31 & 25 & 3 & 3 \\
DeGPT-Qwen & 74 & 69 & 5 & 30 & 27 & 3 & 0 \\
\midrule
\multicolumn{8}{@{}l}{\emph{End-to-end systems}} \\
AutoDecompiler & 7 & 2 & 5 & 2 & 0 & 1 & 1 \\
Nova & 11 & 3 & 8 & 5 & 0 & 3 & 2 \\
SLaDe & 0 & 0 & 0 & 2 & 1 & 1 & 0 \\
GLM-5.2-E2E & 41 & 28 & 13 & 23 & 8 & 11 & 4 \\
\bottomrule
\end{tabular}
\caption{{Differential testing on the GitHub and CVE data of \benchmark{}}, as percentages of each track's denominator $n$. Bld: the candidate compiles; Mat: Matched, matches the reference on every input; Div: Divergence, any divergence on at least one input, including a changed output or bounded observable post-state, an introduced crash, or a hang; C-A: Crash Absence, the validated PoC no longer crashes, defined only on the CVE track and counted separately from Div.}
\label{tab:results-dd}
\end{table}

\paragraph*{LLM-Amplified Divergence in Real Projects}

Table~\ref{tab:results-dd} reports the two tracks with no fixed tests: 291 GitHub library functions and 287 CVE functions. Most systems produce extractable code on 99--100\% of cases; the exceptions are Nova (74\%/81\%), whose non-causal attention mask forbids a fused kernel and forces a dense tensor (100\,GB VRAM for a 40k-token function), and SLaDe (1\%/14\%), whose unmodified source-to-assembly pipeline cannot treat header-dependent library functions as standalone compilation units. The decompilation-tuned end-to-end models also transfer poorly off their curated corpora: Nova builds 58--86\% of the established corpora but only 11\% of GitHub functions (3\% Matched), and AutoDecompiler 8--25\% against 7\% (2\% Matched). GLM-5.2-E2E, reading the same assembly with no decompilation-specific training, reaches 41\% Build and 28\% Matched. This tuned-versus-zero-shot comparison conflates training distribution, scale, and objective. The conflated factors limit this result to an observation about off-distribution robustness.

On real code the refiners buy build rate at a cost to \mbox{behavioral agreement}. LLM4Decompile posts the largest build gain of any refiner, lifting Ghidra from 75\% to 90\%, while its Matched rate falls from 74\% to 62\%. The 28-point gap is mostly changes to the bounded observable post-state (23 points), with 5 points of introduced crashes. Paired per function, 137 functions are Matched under both Ghidra and LLM4Decompile, 77 under Ghidra alone, 42 under LLM4Decompile alone, and 35 under neither (exact two-sided McNemar~\cite{Mcnemar1947NoteOT} $p=.002$): recompilability ranks the refiner higher; behavioral agreement ranks the front end higher. Most of this divergence arises during refinement. Among LLM4Decompile's GitHub functions classified as Divergence, Table~\ref{tab:mechanism} attributes 77 to divergences absent from Ghidra and only 5 to inherited ones. GLM-5.2-Refine (80\%/71\%) and DeGPT-Qwen (74\%/69\%) stay near their front ends (Hex-Rays 81\%/78\%, Ghidra 75\%/74\%), while SK2Decompile cuts Hex-Rays's build rate to 41\% and Idioms builds only 2\%.

\paragraph*{Crash Absence in Vulnerable Code}

On vulnerable code the same rewriting turns into a security problem. CVE functions are harder to recompile because their project-specific types hold even Ghidra and Hex-Rays to 30\% and 23\% Build. LLM4Decompile raises Ghidra's Build to 64\% at 38\% Matched, but 25 of its 183 builds lose the reference crash: a full-track Crash Absence rate of 25/287, or 8.7\% (95\% CI [6.0\%, 12.5\%]), with leave-one-project-out bootstrapping confirming no single project drives it. Raw Ghidra has no Crash Absence cases, and Hex-Rays's three losses arise from mis-striding or UB artifacts. DeGPT-Qwen again tracks Ghidra almost exactly (30\% Build, 27\% Matched, 3\% Divergence, 0\% Crash Absence), whereas GLM-5.2-E2E loses the crash on 11 of 66 builds. The digest covers the bounded observable post-state, so it also flags Divergence when a candidate keeps the crash but changes an output or global. This occurs in 10 of LLM4Decompile's 183 builds and 8 of GLM-5.2-E2E's 66; a crash-only oracle would miss these cases.

The refiners introduce the Crash Absence cases. Attribution is possible only where the front end itself compiles, so we re-score Ghidra and Hex-Rays behind an additive declaration block that supplies the missing placeholder vocabulary while leaving each function body byte-identical. Even then the front end explains little: pooling Divergence and Crash Absence, of the refiners' 114 CVE divergences it had itself diverged on only 27, was Matched on 71, and failed to build on the remaining 16, which are not adjudicable and which Table~\ref{tab:mechanism} folds into \emph{New}, giving 27 and 87. The systems separate sharply. For 64 of LLM4Decompile's 73 divergences, no divergence is inherited from Ghidra. LLM4Decompile expands Ghidra's 85 builds to 183 and incurs 25 Crash Absence cases, whereas DeGPT-Qwen reproduces an already-present Ghidra divergence in 8 of its 9 and has none. More extensive rewriting shifts the failure profile from inherited front-end errors to newly introduced errors.

\subsection{Source-Level Analysis of Divergence}
\label{sec:source-analysis}

\begin{table*}[t]
\centering
\small
\begin{tabular}{@{}l r rr rrr rrr rr rrr@{}}
\toprule
& \multicolumn{9}{c}{Source-level rewriting} & \multicolumn{5}{c}{Divergence attribution} \\
\cmidrule(lr){2-10}\cmidrule(l){11-15}
& & \multicolumn{2}{c}{FE vocab.} & \multicolumn{3}{c}{introduced} & \multicolumn{3}{c}{Fail token} & \multicolumn{2}{c}{GitHub} & \multicolumn{3}{c}{CVE} \\
\cmidrule(lr){3-4}\cmidrule(lr){5-7}\cmidrule(lr){8-10}\cmidrule(lr){11-12}\cmidrule(l){13-15}
System & Pairs & Voc & Scr & Fld & Typ & Cal & $N_F$ & Itr & FE & Inh & New & Inh & New & Fix \\
\midrule
Ghidra & 4170 & 4641 & 0 & 0 & 0 & 0 & 1125 & 0 & 487 & -- & -- & -- & -- & -- \\
Hex-Rays & 4152 & 7884 & 0 & 0 & 0 & 0 & 858 & 1 & 236 & -- & -- & -- & -- & -- \\
\midrule
LLM4Decompile & 3994 & 4358 & 4253 & 255 & 362 & 37 & 504 & 252 & 13 & 5 & 77 & 9 & 64 & 2 \\
SK2Decompile & 3979 & 8096 & 7931 & 843 & 773 & 170 & 668 & 481 & 1 & 1 & 23 & 3 & 11 & 1 \\
Idioms & 3975 & 7584 & 7457 & 241 & 580 & 777 & 870 & 413 & 33 & 0 & 4 & 0 & 1 & 0 \\
GLM-5.2-Refine & 3468 & 6977 & 2768 & 75 & 134 & 202 & 543 & 148 & 81 & 6 & 18 & 7 & 10 & 2 \\
DeGPT-Qwen & 3441 & 3938 & 614 & 0 & 0 & 20 & 904 & 58 & 369 & 5 & 10 & 8 & 1 & 0 \\
\midrule
AutoDecompiler & 4009 & -- & -- & 726 & 1119 & 593 & 2529 & 1050 & 0 & -- & -- & -- & -- & -- \\
Nova & 3668 & -- & -- & 807 & 1172 & 937 & 1237 & 766 & 0 & -- & -- & -- & -- & -- \\
SLaDe & 1987 & -- & -- & 100 & 261 & 36 & 411 & 276 & 0 & -- & -- & -- & -- & -- \\
GLM-5.2-E2E & 3502 & -- & -- & 532 & 604 & 460 & 554 & 282 & 0 & -- & -- & -- & -- & -- \\
\bottomrule
\end{tabular}
\caption{{Source-level rewriting mechanism and divergence attribution, in absolute
counts.} {Source-level rewriting} spans the 40{,}345 pairs of the Pairs column. Voc: front-end vocabulary incidences in the input; Scr: how many the system removes.
Fld, Typ, Cal: introduced fields, types, and callees (Section~\ref{sec:source-analysis}). {Fail token}: $N_F$ is the number of candidates that
failed to compile, of which Itr name a token absent from both input and reference and FE name
front-end vocabulary or an input-only symbol; the remainder cite non-C syntax or misuse of an
existing symbol. {Divergence attribution} pairs each refiner with its own front end on
the same function: of the functions the refiner diverges on, {Inh} counts those where the
front end also diverged and {New} the rest (including unadjudicable functions whose front
end did not build, 27/132 on GitHub, 16/87 on CVE, an upper bound); {Fix} (CVE only) counts front-end Divergences the refiner repairs.}
\label{tab:mechanism}
\end{table*}

Across all function-candidate pairs, we compare each generated output with its corresponding reference and front-end decompiler artifacts, spanning all evaluation axes. This source code and diagnostics level analysis operates without a behavioral comparison oracle, making its dataset a superset of the behaviorally adjudicated corpora. As shown in Table~\ref{tab:mechanism}, we quantify changes to placeholder types, synthetic names, and raw offsets, subsequently identifying the exact token cited by the compiler. A field, type, or callee counts as \emph{introduced} only if its identifier is absent, after comment and string stripping, from both the front-end input and the reference source, with C keywords, standard identifiers, and decompiler vocabulary exempted. A manual inspection of 60 flagged compiled outputs plus 20 unflagged controls found no missed invention; among flags, 12\% of tokens are mere fabrications denoting entities with no counterpart in reference or input, while the rest rename or re-bind real ones, most often substituting a standard-library call for a project macro or wrapper. The divergence association below pools both forms of rewriting. The columns use Scr for removed front-end vocabulary; Fld, Typ, and Cal for introduced fields, types, and callees; and Itr versus FE for failed splices whose diagnostic names an introduced symbol versus a retained front-end artifact. Scr is not applicable to end-to-end inputs, and Idioms' name-canonicalized input makes callee invention part of its task.

Failure mechanisms show patterns that correlate with Divergence rates. Systems whose failures retain front-end artifacts (Ghidra, Hex-Rays, and DeGPT-Qwen) are the only ones to keep Divergence at or below 7.0\% of their GitHub builds, the highest of them being DeGPT-Qwen at 15 of 215; the next-lowest system is GLM-5.2-Refine, at 24 of 232 builds or 10.3\%. Systems that fail because they hallucinate new code (inventions in Table~\ref{tab:mechanism}) perform much worse. Every configuration with GitHub builds where over 40\% of failures involve an introduced symbol exhibits a Divergence rate of 20.3\% or higher, the lowest being SK2Decompile at 24 of 118 builds. The association also holds within systems. Among a system's own compiled candidates, functions whose output carries an introduced field, type, or callee diverge far more often than its invention-free functions: 53\% versus 20\% pooled on GitHub (odds ratio 4.6, Fisher $p<10^{-6}$) and 59\% versus 31\% on the CVE track ($p<10^{-4}$). The gap persists within function-length terciles. Shared drivers such as function difficulty and type richness may still contribute, so the association supports the proposed mechanism without establishing causality. The reported gap is conservative because Divergence rates cover only code that compiles; many hallucinated programs fail to build and are excluded.

The main pattern is a migration from visible unknowns in symbolic decompilers to confident inventions. Among recompilation failures, 43\% for raw Ghidra and 28\% for raw Hex-Rays name a retained front-end artifact such as \texttt{DAT\_*} or \texttt{undefined4}. Only one Hex-Rays failure names an introduced symbol, and the remaining failures involve non-C syntax or misuse of a real symbol. With the exception of DeGPT-Qwen, rewriting systems remove 40--98\% of the front-end vocabulary. Turning \texttt{*(\_DWORD*)(a1+40)} into C requires the model to supply a field, type, or callee absent from the binary. Introduced symbols appear in 42--72\% of failed splices for the tuned refiners and end-to-end models. Even GLM-5.2-Refine, the most conservative rewriter by this measure, reaches 27\%. Inventions that survive compilation lead outputs to diverge, yielding either Divergence or Crash Absence. DeGPT-Qwen edits pseudocode in place, removes 16\% of the vocabulary, and stays at or below 1\% on each invention axis. Its failure profile consequently resembles Ghidra's: 41\% front-end tokens compared with 43\% for raw Ghidra.

Analysis on the decompiler outputs shows that SK2Decompile rewrites aggressively, including target renames, storage-class and address-of changes, and introduced fields; its real-world build rate is roughly half that of Hex-Rays. LLM4Decompile preserves most fields and callees but makes small, compilable inventions such as \texttt{u32}, \texttt{Mat4}, or a changed hexadecimal constant, which tend to surface as Divergence. DeGPT-Qwen edits conservatively and remains close to Ghidra on both vocabulary removal and invention; full per-system fingerprints appear in the supplement. The same behavior appears at different stages of the pipeline. Curated corpora mostly contain flat scalar signatures, leaving little room to invent fields; introduced-field prevalence stays at or below 2\%. Most outputs still compile, so incorrect inventions appear as Divergence. Real code uses project-specific structs and types. On CVE functions, introduced-field prevalence reaches 56--72\% for SK2Decompile, Nova, and GLM-5.2-E2E. Some inventions prevent compilation and contribute to the Build collapse in Tables~\ref{tab:results} and~\ref{tab:results-dd}. Those that compile result in Divergence when values change, or result in Crash Absence when the vulnerability depends on the introduced type width, constant, or guard.

\section{Conclusion}

\benchmark{} builds a behavioral comparison oracle from fuzzer-generated inputs derived from the reference function, together with a precise splicing procedure, extending evaluation of behavioral agreement to functions without released tests. Across real and vulnerable code, the results show that recompilability and re-executability fail to capture behavioral divergence, and source analysis connects that divergence to a shift from visible front-end unknowns to introduced tokens.

\section{Acknowledgments}

This work was supported by NSF award CCF-2316159. 
This work was also supported in part through computational resources provided by Syracuse University. The authors gratefully acknowledge use of the OrangeGrid / HTC Campus Grid, supported by NSF award ACI-1341006, and technical support from Syracuse University’s Cyberinfrastructure Engineer, supported by NSF award ACI-1541396.

\bibliography{aaai2027}

\end{document}